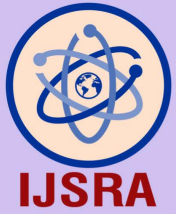

International Journal of Science and Research Archive

eISSN: 2582-8185
Cross Ref DOI: 10.30574/ijsra
Journal homepage: https://ijsra.net/

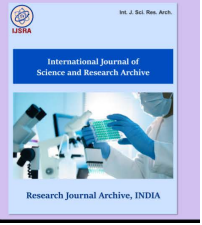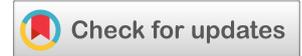(Review Article)

# A comprehensive survey of cybercrimes in India over the last decade

Sudhanshu Sekhar Tripathy *

*Department of Computer Science and Engineering, C.V Raman Global University Bhubaneswar Odisha, India.*International Journal of Science and Research Archive, 2024, 13(01), 2360–2374

Publication history: Received on 31 August 2024; revised on 10 October 2024; accepted on 12 October 2024

Article DOI: https://doi.org/10.30574/ijsra.2024.13.1.1919**Abstract**

Since the 1990s, the integration of technology into daily life has led to the creation of an extensive network of interconnected devices, transforming how individuals and organizations operate. However, this digital transformation has also spurred the rise of cybercrime, criminal activities perpetrated through networks or computer systems. Cybercrime has become a global concern, presenting significant challenges to security systems. Although advancements in digital technology have enhanced efficiency, they have also opened new avenues for exploitation by cybercriminals, highlighting the urgent need for advanced cybersecurity measures. The escalating number of cyberattacks and associated risks in the past decade highlights the critical importance of protecting sensitive data and safeguarding information systems. Cybercrimes range from financial fraud and phishing scams to identity theft and online harassment, posing substantial risks to both individuals and organizations. In response, governments, law enforcement agencies, and cybersecurity units have intensified their efforts to address these threats. In recent years, India has experienced a significant surge in cybercrime incidents, with a notable increase in cases involving ransomware, data breaches, and social engineering attacks. The growing penetration of internet services, the expansion of e-commerce, and the rapid adoption of digital payment systems have made individuals and organizations more vulnerable to cyber threats. Key areas affected include banking, healthcare, and government sectors, which are frequently targeted due to the sensitive nature of the data they handle. To combat these risks, there is an increasing focus on public awareness, cybersecurity education, and robust regulatory frameworks. This paper provides an in-depth analysis of cybercrime, with a focus on developing innovative prevention strategies, strengthening internal security protocols, and classifying key cybercrime terminologies to better understand their implications for digital infrastructure.

**Keywords:** Cybercrime; IT act 2000; Fraud Cases; Cybersecurity; Cyberattack## 1. Introduction

The rapid advancement of digital communication technologies has redefined the global landscape of information exchange, but this progression has also fueled a surge in cybercrime. Cybercrime refers to unlawful activities carried out through or targeting digital infrastructures, with information systems functioning as either the instrument or the victim of these offenses. Such crimes encompass a broad spectrum, including phishing, identity theft, financial fraud, cyberbullying, the distribution of illicit content, denial-of-service (DoS) attacks, spamming, and the trafficking of illegal goods through online channels. Technically, cybercrime is distinguished from general internet-related offenses by its reliance on digital networks whether public or private and organizational information systems as tools or targets. Common manifestations of cybercrime include sophisticated phishing campaigns, unauthorized access to financial data, credit card fraud, corporate espionage, the dissemination of child exploitation material, and the coordination of cyberterrorist activities. A recurring theme across these offenses is the illegal access to, and exploitation of sensitive or confidential information. In response to the global escalation of cybercrime, many governments have instituted stringent legal frameworks and regulatory mechanisms aimed at reducing the incidence of such crimes. Despite these efforts, the dynamic and ever-evolving nature of information technology continues to expose new vulnerabilities, offering cybercriminals fresh avenues for exploitation [1].

* Corresponding author: Sudhanshu Sekhar TripathyCopyright © 2024 Author(s) retain the copyright of this article. This article is published under the terms of the Creative Commons Attribution Liscense 4.0.

As a result, ensuring robust cybersecurity measures has become a priority for researchers, policymakers, and law enforcement agencies. This study endeavors to provide a detailed exploration of the various forms and contexts of cybercrime, alongside an analysis of the underlying factors that facilitate these activities. Through an examination of real-world case studies and the continually evolving methodologies employed by cybercriminals, this research aims to contribute to the development of advanced defense mechanisms. The insights presented are anticipated to inform the future direction of cybersecurity strategies and aid in the mitigation of cybercrime incidents globally.

## 2. Evolution of Cybercrime in India (2014–2024)

By 2024, approximately 5.45 billion individuals, representing 67.1% of the global population, are projected to have internet access. Asia, driven primarily by East and South Asian nations, will account for the largest share of this user base, with nearly 2.9 billion users. Northern Europe is expected to achieve the highest internet penetration rate, reaching approximately 97.5%, reflecting the region's advanced digital infrastructure and widespread accessibility. Conversely, regions such as the Middle East and Africa exhibit significantly lower internet penetration rates. This disparity can be attributed to several factors, including underdeveloped digital infrastructure, limited technological resources, economic constraints, and a lack of digital literacy. These challenges hinder the widespread adoption of internet services in these regions, exacerbating the global digital divide. Addressing these issues remains critical for promoting equitable access to information and participation in the global digital economy.

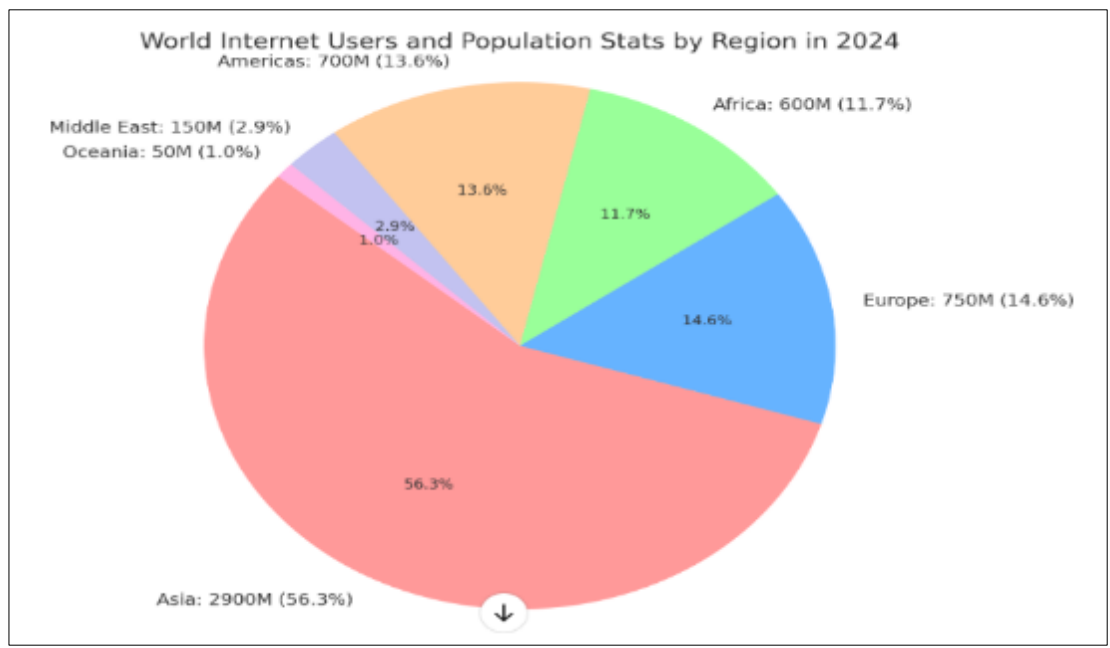

**Figure 1** Internet users by world regions-2024 [2]

As depicted in figure 1, the pie chart provides a detailed breakdown of the projected number of internet users by region in 2024, including both the total number of users in millions and their respective percentage shares of the global internet population. This representation enables a comprehensive visualization of the regional distribution of internet usage, offering a clear and precise overview of the global landscape. The format enhances the clarity of the data, facilitating an in-depth understanding of how internet access is geographically distributed across various regions. The 2024 pie chart illustrates the global distribution of internet users by region. Asia holds the largest share, accounting for 56.3% (2.9 billion users), driven by its vast population and increasing internet access. Europe follows with 14.6% (750 million users), showcasing strong connectivity. The Americas contribute 13.6% (700 million users), while Africa represents 11.7% (600 million users), reflecting rapid growth in internet usage. The Middle East and Oceania have smaller shares, at 2.9% (150 million users) and 1.0% (50 million users), respectively. This distribution highlights disparities in internet penetration and accessibility across different regions.

**Table 1** World Internet Usage and Population Statistics (August 30, 2024)

| Region | Internet Penetration (%) | Population (Millions) | Internet Users (Millions) |
|---|---|---|---|
| Northern Europe | 97.5% | 108.2 | 105.4 |
| Northern America | 96.9% | 375.5 | 363.8 |
| Western Europe | 94.5% | 196.2 | 185.3 |
| Southern Europe | 90.2% | 152.7 | 137.6 |
| Eastern Europe | 88.4% | 293.0 | 259.0 |
| Southern America | 82.6% | 434.4 | 358.8 |
| Central America | 78.4% | 182.5 | 143.0 |
| Eastern Asia | 77.7% | 1,678.4 | 1,303.4 |
| Oceania | 77.4% | 43.6 | 33.7 |
| Central Asia | 76.5% | 74.9 | 57.3 |
| Southern Africa | 75.9% | 68.9 | 52.3 |
| Western Asia | 75.3% | 277.3 | 208.7 |
| South-Eastern Asia | 74.0% | 672.3 | 497.4 |
| Caribbean | 70.3% | 44.4 | 31.2 |
| Northern Africa | 70.1% | 253.7 | 177.7 |
| Southern Asia | 53.6% | 1,972.1 | 1,056.0 |
| Western Africa | 42.3% | 426.5 | 180.2 |
| Middle Africa | 32.2% | 200.3 | 64.5 |
| Eastern Africa | 26.8% | 473.7 | 126.9 |

As of August 30, 2024, the table 1, provides insights into internet penetration rates, population figures, and internet users across various global regions. Northern Europe leads with a penetration rate of 97.5%, indicating near-universal access among its 108.2 million population, with 105.4 million users. Northern America follows closely at 96.9%, with 363.8 million users out of a population of 375.5 million. Western Europe also exhibits high connectivity, achieving a 94.5% penetration rate, with 185.3 million users from 196.2 million people. In contrast, Southern Asia shows a significantly lower penetration rate of 53.6%, with over 1 billion users among nearly 2 billion people, highlighting challenges in access. Middle Africa (32.2%) and Eastern Africa (26.8%) report even lower penetration rates, indicating substantial barriers to connectivity. Moderate to good penetration rates are observed in regions like Southern America (82.6%) and Eastern Asia (77.7%), reflecting varying levels of internet access. This data underscores the digital divide, showcasing regions with high connectivity and those facing significant obstacles. Understanding these statistics is essential for policymakers focused on improving internet access and fostering digital inclusion on a global scale.

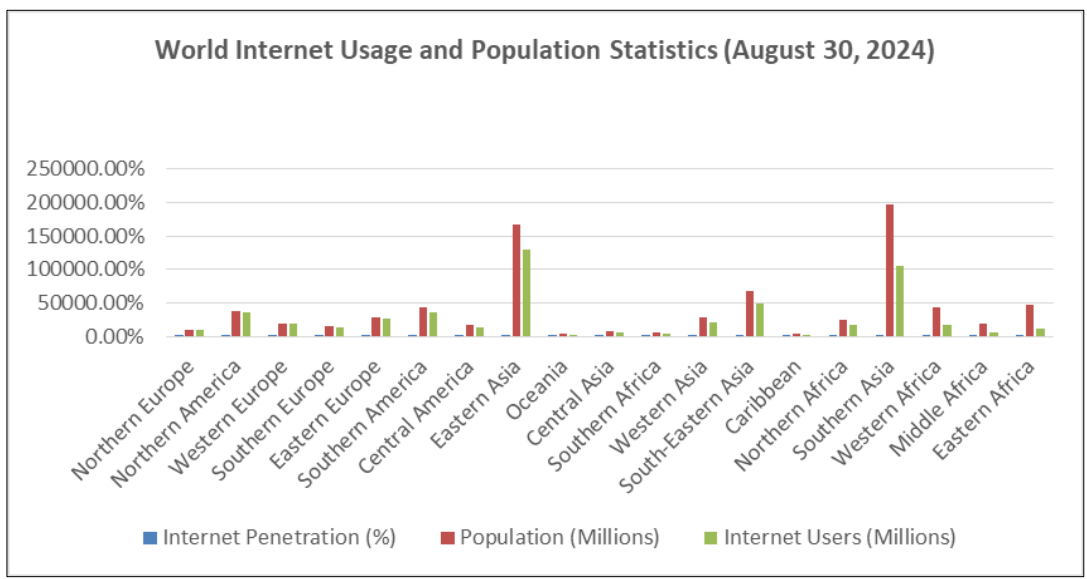

**Figure 2** World Internet Usage and Population Statistics (August 30, 2024)

Figure 2 illustrates the pronounced global disparities in internet access, with developed regions such as Northern Europe and Northern America exhibiting the highest penetration rates. In contrast, regions like Middle and Eastern Africa display significantly lower levels of internet connectivity. These differences underscore the ongoing digital divide, where developed areas benefit from advanced infrastructure and widespread access, while less developed regions face substantial barriers to digital inclusion due to economic, infrastructural, and technological challenges.

**Table 2** World Internet Usage and Population Statistics (2014-2024)

| Year | World Population (Billion) | Internet Users (Billion) | Internet Penetration Rate (%) |
|---|---|---|---|
| 2014 | 7.2 | 2.8 | 39% |
| 2015 | 7.3 | 3.2 | 43% |
| 2016 | 7.4 | 3.5 | 47% |
| 2017 | 7.5 | 3.7 | 49% |
| 2018 | 7.6 | 4.0 | 52% |
| 2019 | 7.7 | 4.3 | 56% |
| 2020 | 7.8 | 4.6 | 59% |
| 2021 | 7.9 | 4.9 | 62% |
| 2022 | 8.0 | 5.2 | 65% |
| 2023 | 8.1 | 5.4 | 67% |
| 2024* | 8.2 | 5.6 | 68% |

Between 2014 and 2024, the global internet user base witnessed significant expansion, closely aligned with overall population growth. In 2014, the global population was approximately 7.2 billion, with only 2.8 billion individuals connected to the internet, resulting in a penetration rate of 39%. Over the subsequent decade, this number increased substantially, with 5.6 billion users projected by 2024, bringing the penetration rate to 68%. This substantial growth is attributable to several factors, including technological advancements, the widespread adoption of smartphones, and the ongoing development of internet infrastructure. The increased connectivity has had transformative effects on multiple sectors, reshaping communication, commerce, and education by facilitating greater access to information and digital services. However, despite this overall progress, significant regional disparities in internet access remain, particularly in less developed areas where infrastructure and economic challenges continue to limit digital inclusion.

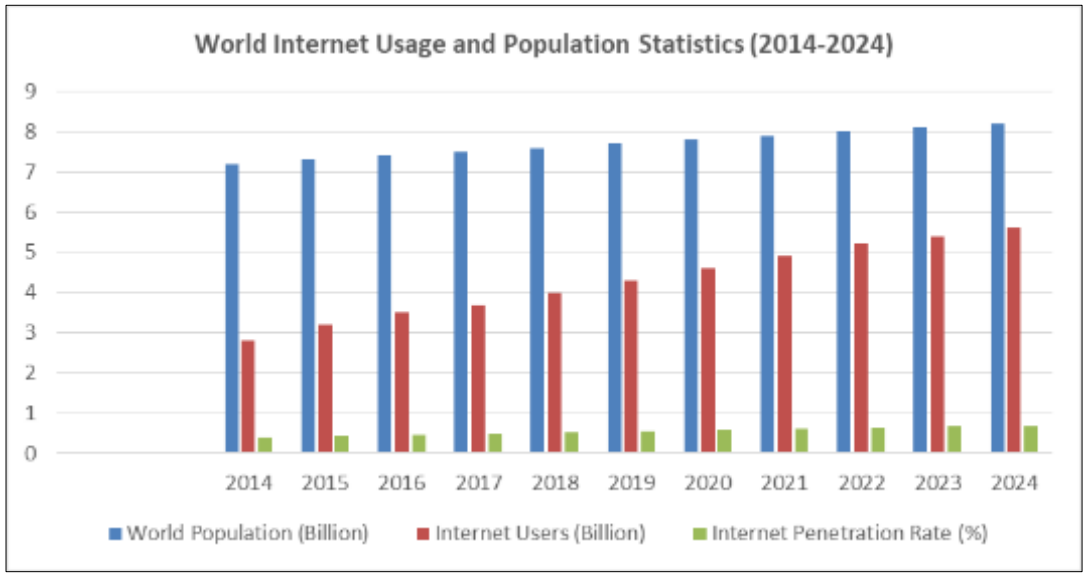

**Figure 3** World Internet Usage and Population Statistics (2014-2024)

Figure 3, titled "World Internet Usage and Population Statistics (2014-2024)," presents a bar graph that visually demonstrates the parallel growth of internet users and the global population over the ten-year period. The blue bars represent global population figures, showing a steady increase from 7.2 billion in 2014 to an estimated 8.2 billion by 2024. In contrast, the red bars depict the expansion of the internet user base, which has grown significantly from 2.8 billion in 2014 to 5.6 billion in 2024. The green line tracks the internet penetration rate, which exhibits a notable increase from 39% to 68% over the same timeframe.

This visual representation highlights the increasing significance of internet connectivity as a fundamental component of global communication and access to information. It reflects not only advancements in digital infrastructure but also the expanding role of the internet in shaping economic, social, and educational systems worldwide.

## 3. Understanding the Landscape of Cybercrime: Key Definitions and Concepts

- Cyber Crime: Illegal activities conducted online, including hacking, identity theft, and online fraud that exploit computer networks and devices for malicious purposes.
- Internet: The internet is a global network enabling communication, web access, and data transfer, transforming societal and economic interactions through email, file sharing, and real-time collaboration across diverse sectors worldwide.
- Web Browser: A web browser is a software application designed to retrieve, display, and interact with information resources hosted on World Wide Web servers. These resources, such as web pages, images, and multimedia content, are accessed through a Uniform Resource Locator (URL), which serves as the address that uniquely identifies the desired server and resource. Web browsers enable seamless navigation and retrieval of web-based content, playing a pivotal role in facilitating access to digital information and enhancing user interaction with the vast repositories of knowledge available online.
- ISP: An Internet Service Provider (ISP) offers internet access to individuals and organizations, enabling connectivity through various technologies. ISPs play a critical role in managing online traffic and ensuring reliable internet services
- IP address: An IP address (Internet Protocol address) is a unique numerical identifier assigned to each device on a network, facilitating communication and data exchange. For example, 152.58.166.204 denotes a specific device.
- IP address spoofing: IP address spoofing involves the forgery of a user's IP address, allowing an intruder to impersonate the legitimate user. This technique alters packet headers, making malicious activities appear to originate from a trusted source.
- Email spoofing: Email spoofing is when a hacker forges an email address to make it appear as if it's from a trusted sender. This technique is used for malicious activities, such as phishing or spreading malware.
- Computer virus: A computer virus is a malicious program that infects a user's computer without consent, capable of damaging files and replicating itself across systems. These viruses often disrupt normal operations

- and compromise data integrity.
- Spyware: Spyware is malicious software secretly installed on a user's device to collect personal information, such as browsing habits or keystrokes. Examples include keyloggers, which capture every keystroke made.
- Computer Worm: A computer worm is a self-replicating program that spreads to other computers over a network, exploiting security vulnerabilities. Unlike viruses, worms do not attach to existing files, allowing them to propagate independently.

**3.1. Comprehensive Overview of Cyber Crime Terminologies in India:**

- Hacking: Unauthorized access to computer systems or networks, often aimed at stealing data or disrupting services, which is a punishable offense under the Information Technology Act in India.
- Phishing: Fraudulent attempts to acquire sensitive information, such as passwords or bank details, by impersonating a trustworthy entity in emails or messages, often targeting unsuspecting individuals.

*3.1.1. Example of Phishing*

- **Scenario:** Here's an original example of a phishing email targeting bank customer:
- **Subject:** Important: Verify Your Account Immediately!
- **From:** service@sbi.com

Dear [Manoranjan Sahoo],

We noticed some unusual activity on your account and need your immediate attention to ensure your security. Please verify your account information by clicking the link below:

*3.1.2. Verify My Account*

http://sbi-update-confirmation.com

Failure to verify your account within 24 hours will result in temporary suspension for security purposes.

Thank you for your cooperation.

Sincerely,
**Customer Support Team** [State Bank Of India]

- Identity Theft: The illegal use of someone's personal information, like Aadhaar numbers or bank account details, to commit fraud or impersonate them, resulting in financial and emotional consequences.
- Ransomware: Malicious software that encrypts a victim's files, demanding a ransom for their decryption, increasingly targeting individuals and organizations in India, causing severe operational disruptions.
- Cyberstalking: The use of digital platforms to harass or intimidate individuals, often leading to fear or anxiety, and can involve tracking online activities and sending threatening messages
- Online Fraud: Deceptive schemes executed over the internet, such as fake job offers or investment scams, designed to exploit victims for financial gain, prevalent in various online platforms.
- Denial of Service (DoS) Attack: An assault aimed at making a service unavailable by overwhelming it with traffic or exploiting vulnerabilities, hindering legitimate users from accessing essential online resources.
- Data Breach: Incidents where unauthorized individuals gain access to confidential data, exposing sensitive information of individuals or organizations, and leading to significant financial and reputational damage.
- Digital Harassment: Harassment conducted through online platforms, such as social media, impacting the victim's mental well-being and safety, and often involving threatening messages or unwanted contact.
- Cyber Bullying: Cyberbullying is the act of harassing or humiliating someone online through social media, messaging apps, or other digital platforms, commonly affecting teenagers in India.
- ATM card Fraud: Involves skimming devices or hacking methods to steal credit or debit card information at ATMs. It can also involve phishing attempts to trick users into providing their PIN.
- Cryptojacking: This form of cybercrime involves unauthorized use of someone's computer to mine cryptocurrencies. India has seen a rise in cryptojacking incidents as cryptocurrency adoption grows.
- Vishing: Vishing (voice phishing) involves cybercriminals making phone calls to trick people into giving sensitive personal information, such as banking details or OTPs.
- SIM Swap Fraud: In this type of fraud, attackers swap a victim's SIM card to gain access to their bank accounts or social media by intercepting two-factor authentication codes.
- Child Exploitation Online: This refers to the exploitation of minors on digital platforms through child

- pornography, grooming, or solicitation, often carried out through social media.
- Botnet Attack: Botnets are networks of compromised computers controlled by cybercriminals. They are used to conduct Distributed Denial of Service (DDoS) attacks or steal data.
- Spoofing: Spoofing involves disguising communication from an unknown source as being from a known, trusted source. Email spoofing and IP spoofing are common techniques used in India.
- Dark Web Crimes: The dark web is used for illicit activities, including the sale of drugs, weapons, stolen data, and illegal services. Many cybercriminals in India operate on the dark web for anonymity.
- Social Engineering: Social engineering manipulates individuals into divulging confidential or personal information. Techniques like phishing, pretexting, and baiting are examples of this crime.
- Deepfake Technology: Deepfakes use AI to create highly realistic but fake images, videos, or audio clips. In India, deepfakes have been used for misinformation or to create fake content for extortion.
- Cyber Espionage: Cyber espionage refers to the illegal exploitation of computer systems and networks to gain access to confidential information, often for political or economic gain.
- Doxxing: Doxxing involves the public release of an individual's personal information (such as home address, phone number, etc.) to harass or harm them.
- Betting Fraud Scam: Betting fraud scams deceive victims through fake betting platforms or manipulated odds, leading to financial loss. Scammers often promise guaranteed wins or insider tips to lure individuals into fraudulent schemes.
- Electricity Bill Scam: Electricity bill scams involve fraudulent schemes where scammers impersonate utility companies, claiming unpaid bills to extort money. Victims may receive threatening calls or emails demanding immediate payment to avoid disconnection.
- Loan Lottery Scam: Loan lottery scams trick victims into believing they've won a lottery for a loan. Scammers typically ask for fees or personal information to claim the nonexistent prize, leading to financial loss.
- Honey Trap Scam: A honey trap scam involves using romantic deception to manipulate victims into divulging personal information or sending money. Scammers often create fake profiles on dating sites to establish trust and exploit emotions.
- Digital Arrest: Digital Arrest in cybercrime refers to the process of apprehending individuals engaged in online illegal activities, such as hacking, identity theft, and fraud. This involves law enforcement investigating crimes, collecting digital evidence, and seizing electronic devices while following legal protocols to ensure a thorough and lawful resolution of cases.
- Banking Scam: A banking scam involves fraudulent schemes aimed at deceiving individuals or financial institutions to steal money, sensitive information, or gain unauthorized access to accounts. These scams can take the form of phishing emails, fake calls from supposed bank representatives, unauthorized transactions, or deceptive loan offers. Cybercriminals use various tactics to manipulate victims into revealing account details or transferring funds, leading to significant financial losses for both individuals and banks.
- OLX Frauds: OLX frauds typically involve scammers posing as buyers or sellers on the online marketplace to deceive users. Common tactics include fake payment confirmations, requests for advance payments, or phishing links that steal personal data. Scammers often exploit the trust of users by offering deals too good to be true, leading to financial losses for unsuspecting victims. Always verify the authenticity of buyers or sellers before making transactions.
- Online Dating Scam: Online dating app scams involve fraudsters creating fake profiles to exploit users emotionally and financially. Scammers often present themselves as ideal matches to gain trust before requesting money for fabricated emergencies or offering enticing but non-existent deals. Victims may also fall for phishing attempts, revealing personal information. Awareness and caution are vital; users should verify profiles and never share sensitive information.

### 3.2. Effective Strategies to Prevent Cyber Fraud Scams

- Use multi-factor authentication (MFA) for secure access.
- Encrypt data during storage and transmission to protect it.
- Regularly update software and apply security patches.
- Deploy intrusion detection systems (IDS) to monitor network activities.
- Conduct cybersecurity training to raise awareness about threats like phishing.
- Maintain automatic backups to recover data in case of attacks.
- Install and update antivirus and anti-malware software.
- Implement zero-trust security to continuously verify users and devices.
- Monitor network traffic for anomalies and suspicious behavior.
- Share threat intelligence with other organizations to stay informed about new risks.

## 3.3. Fundamental Cybersecurity Guidelines Proposed by the Information Technology Promotion Agency

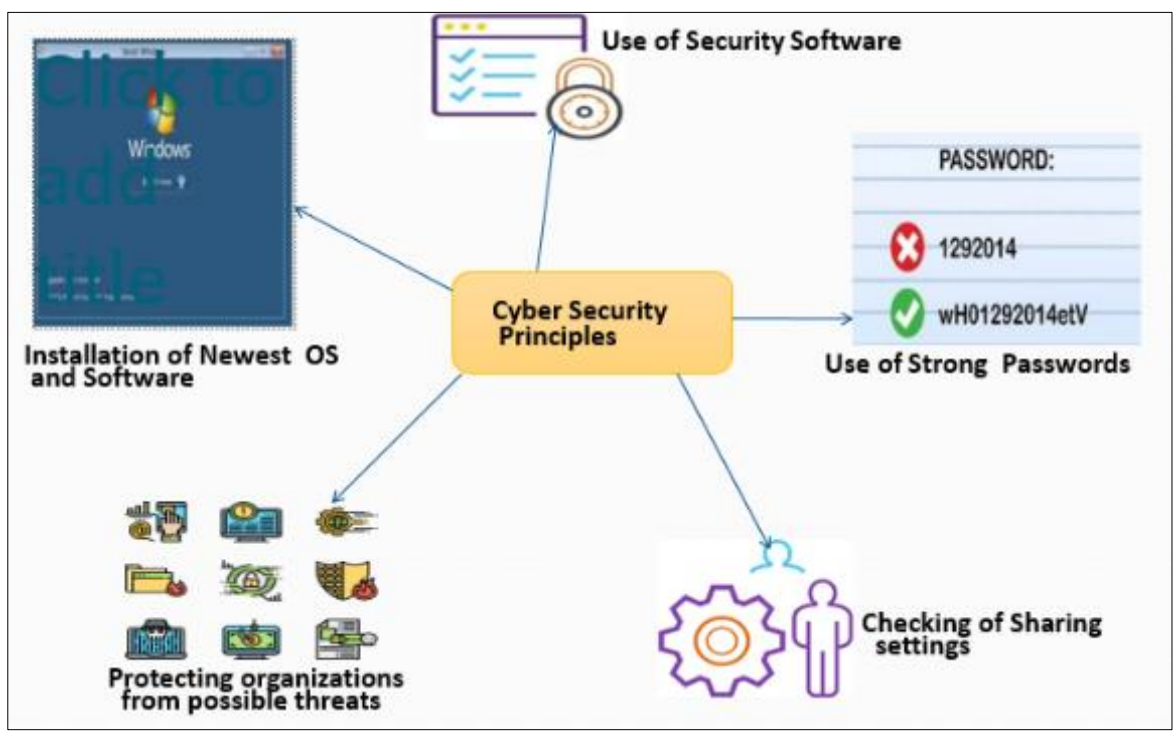

**Figure 4** Cybersecurity Guidelines Proposed by the Information Technology Promotion Agency

*3.3.1. Cybersecurity is essential in safeguarding organizations against potential threats. Key principles include:*

- Installation of the Latest Operating Systems and Software: Regularly updating systems and applications ensures vulnerabilities are patched, reducing the risk of cyberattacks.
- Use of Security Software: Implementing robust security software is crucial for detecting and preventing malware and unauthorized access.
- Adoption of Strong Passwords: Strong, unique passwords are vital for protecting accounts. Users should avoid easily guessable passwords and consider using password managers for secure storage.
- Checking Sharing Settings: Regularly reviewing sharing settings on networks and devices can help prevent unauthorized access to sensitive information.

By adhering to these principles, organizations can enhance their cybersecurity posture and better protect their digital assets.

## 4. An Overview of Cyber Crime Cases in India

Between 2014 and 2024, India has experienced a marked increase in cybercrime incidents, influenced by rapid technological advancement and digital adoption. Key developments during this period include:

- Cybercrime Surge: India has indeed seen a marked increase in cyber-crimes, with a significant rise from 9,622 cases in 2014 to over 77,000 in 2024 (till August). This trend reflects growing digital adoption, vulnerabilities in cybersecurity, and the expanded attack surface due to new technologies like cloud computing and IoT.
- Government Initiatives: The Indian government has taken various measures to address the growing cyber threats, including the establishment of the Indian Cyber Crime Coordination Centre (I4C) in 2020. This initiative works as a central hub for handling cybercrime incidents and enhancing law enforcement capabilities. The National Cyber Crime Reporting Portal also enables individuals to report crimes, with a focus on crimes against women and children.
- Legal Framework: The Information Technology Act (2000) has been amended to address the evolving threat landscape, but enforcement remains a challenge. The legal framework has been bolstered by other government initiatives like CERT-In, which issues advisories to improve cybersecurity across sectors
- Sector-Specific Threats: Sectors like finance, healthcare, and education faced unique challenges, prompting a need for enhanced security measures. (The New Indian Express) [8]

- Impact on Businesses: Highlight the financial implications of cybercrime on businesses, including loss of revenue, damage to reputation, and the costs associated with data breaches and recovery efforts.
- Emerging Threats: Mention emerging forms of cybercrime, such as ransomware attacks, which have gained prominence, targeting organizations and demanding hefty ransoms for data recovery.
- Cybersecurity Investments: Note the increasing investments in cybersecurity by both the government and private sectors, aimed at developing advanced technologies and strategies to prevent cyber threats.
- Public-Private Partnerships: Emphasize the importance of collaboration between the government, law enforcement agencies, and private companies to share intelligence and resources to combat cybercrime effectively.

These efforts highlight the ongoing battle against cybercrime, emphasizing the importance of collaborative strategies between government, law enforcement, and citizens to safeguard digital infrastructure in India.

**Table 3** Cyber Crimes/Cases Registered and Persons Arrested under the IT Act during 2014 - 2024

| Year | Cases Registered | Persons Arrested |
|---|---|---|
| 2014 | 9,622 | 5,752 |
| 2015 | 11,592 | 8,121 |
| 2016 | 12,317 | 8,613 |
| 2017 | 21,796 | 9,622 |
| 2018 | 27,248 | 18,930 |
| 2019 | 44,546 | 21,796 |
| 2020 | 50,035 | 24,064 |
| 2021 | 52,974 | 25,789 |
| 2022 | 65,893 | 27,612 |
| 2023 | 75,656 | 34,597 |
| 2024* Till August | 77,858 | 36,235 |

The data from 2014 to 2024, table 3, illustrates a significant increase in cybercrime cases and arrests, highlighting the urgency of addressing this growing threat. In 2014, there were 9,622 cases reported, which surged to 77,858 by August 2024. This alarming rise reflects the increasing sophistication and frequency of cybercriminal activities. Similarly, the number of arrests has also grown, from 5,752 in 2014 to 36,235 in August 2024. While law enforcement agencies are making efforts to apprehend offenders, the sharp increase in cases indicates that the problem is escalating faster than authorities can respond. This trend underscores the necessity for robust cybersecurity measures and proactive public awareness campaigns to equip individuals and organizations with the knowledge and tools to protect themselves. As cyber threats continue to evolve, a coordinated effort between law enforcement, policymakers, and the community is essential to combat this pervasive issue effectively.

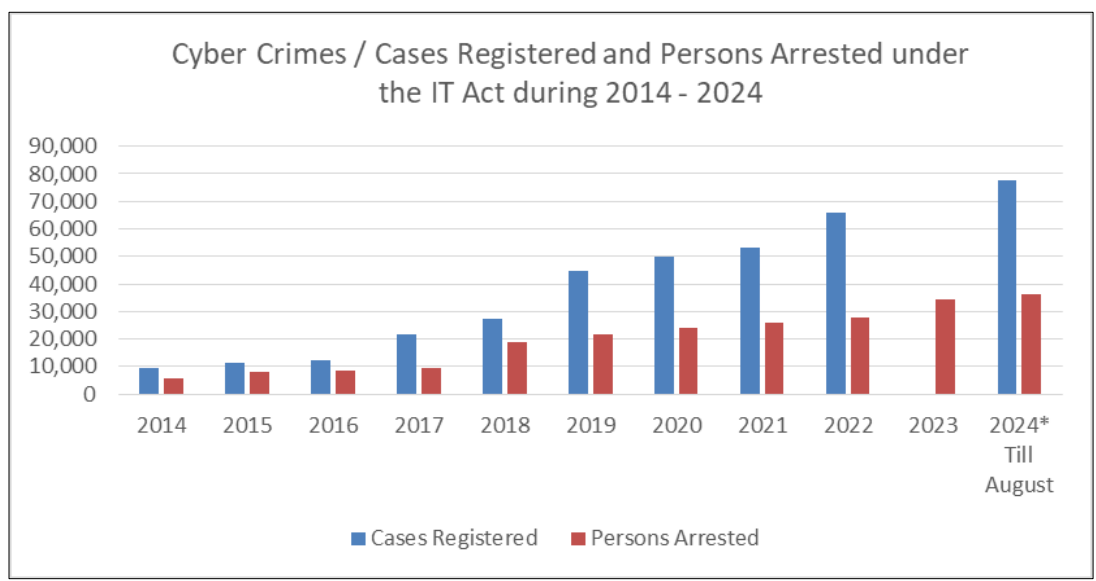

**Figure 5** Cyber Crimes/Cases Registered and Persons Arrested under the IT Act during 2014 – 2024

Figure 5 illustrates a substantial escalation in the total number of cybercrimes reported and the corresponding arrests made between 2014 and 2024. This upward trajectory underscores the increasing prevalence of cybercriminal activities and the evolving response of law enforcement agencies. The data reflect heightened awareness and concern regarding cybersecurity threats, indicating that authorities are intensifying their efforts to address and mitigate the impacts of cybercrime in this dynamic landscape.

### 4.1. Cybercrime Surge in India (2014-2024): A Decade of Digital Threats and Legal Interventions

Over the decade from 2014 to 2024, India has witnessed a dramatic surge in cybercrime, reflecting both the country's rapid digital transformation and the corresponding rise in cyber threats. As more citizens and businesses embrace digital technologies, vulnerabilities in cybersecurity have also become more apparent, leading to a sharp increase in the number of reported cyber incidents. In 2014, there were only around 9,622 cases of cybercrime, but by the end of 2024, this number is projected to exceed 80,000, illustrating the severity of the issue. Key drivers of this increase include the widespread adoption of internet services, mobile phones, and digital banking systems. This expansion of digital infrastructure, while beneficial for economic growth, has also provided opportunities for cybercriminals. Common forms of cybercrime during this period have included identity theft, ransomware attacks, data breaches, phishing schemes, and online fraud.

In response, the Indian government has taken several important steps to combat cybercrime. One of the most significant interventions has been the establishment of the Indian Cyber Crime Coordination Centre (I4C), which serves as a hub for coordinating responses to cyber threats. The National Cyber Crime Reporting Portal has also been introduced to enable citizens to report cyber incidents directly, with a particular focus on crimes against women and children. To strengthen the legal framework, the Information Technology Act (2000) has been amended multiple times to better address the complexity of modern cybercrimes. Law enforcement agencies have been trained to handle digital crimes more effectively, but challenges remain in terms of cybersecurity awareness and resources across the country. The growing number of attacks targeting critical sectors such as finance, healthcare, and government highlights the urgent need for stronger cybersecurity measures. Collaborative efforts between the government, businesses, and individuals are crucial to building a resilient digital ecosystem in India. As the country continues to embrace new technologies, ensuring robust cyber defenses will be key to safeguarding its digital future.

The Information Technology Act, 2000, was enacted to address crimes that arise from technological advancements, establishing a legal framework to combat the evolving landscape of cybercrimes. It specifically targets offenses such as hacking, data theft, and online fraud, providing necessary provisions for digital transactions and electronic records. Conversely, the Indian Penal Code (IPC) is a comprehensive legal document that encompasses traditional crimes like theft, assault, and fraud, which can also apply to cyber-related offenses when they occur in a digital context. Together, these laws ensure a robust legal approach for prosecuting offenders and delivering justice in cybercrime cases.

**4.2. Cybersecurity Investment Trends: CAGR Insights for 2014-2026:**

The cybersecurity market has experienced substantial growth from 2014 to 2026, driven by an increasing frequency of cyber threats and a heightened awareness of the importance of data protection. In 2014, the global cybersecurity market was valued at approximately **$75 billion**. By 2026, this figure is projected to surpass **$366 billion,** reflecting a Compound Annual Growth Rate (CAGR) of around **12.5%. [13] [14].** Several factors contribute to this impressive growth. The rise in digital transformation across various industries, including finance, healthcare, and retail, has led to an expanded attack surface for cybercriminals. This shift has prompted organizations to invest significantly in cybersecurity solutions to safeguard their sensitive data. Additionally, the COVID-19 pandemic accelerated remote work trends, further underscoring the need for robust cybersecurity measures as employees accessed corporate networks from various locations.

The cybersecurity landscape is also evolving with the integration of advanced technologies such as artificial intelligence and machine learning, which enhance threat detection and response capabilities. Moreover, regulatory frameworks like the General Data Protection Regulation (GDPR) and the California Consumer Privacy Act (CCPA) have compelled organizations to adopt stringent security protocols. In brief, the CAGR of approximately **12.5%** from 2014 to 2026 highlights the critical role of cybersecurity in today's digital world. As cyber threats continue to grow in complexity and scale, businesses must prioritize investments in advanced security solutions to protect their operations and maintain compliance with regulatory requirements.

**Table 4** Cybersecurity Investment Trends: CAGR Insights for 2014-2026

| Year | Market Value (Billion USD) |
|---|---|
| 2014 | 75 |
| 2016 | 100 |
| 2018 | 152 |
| 2020 | 200 |
| 2022 | 245 |
| 2024 | 300 |
| 2026 | 366 |

Table 4. presents the projected growth of the global cybersecurity market from 2014 to 2026. Starting at $75 billion in 2014, the market shows a consistent upward trend, driven by increasing cyber threats and a growing focus on data protection. By 2016, the market had grown to $100 billion, and it reached $152 billion by 2018. The rapid expansion continued through 2020, with a value of $200 billion, fuelled by digital transformation and remote work trends. In 2022, the market further escalated to $245 billion, with expectations of hitting $300 billion in 2024. By 2026, the market is projected to surpass $366 billion, reflecting the persistent need for advanced cybersecurity solutions across industries. This growth highlights the sector's critical role in safeguarding against increasingly sophisticated cyber threats.

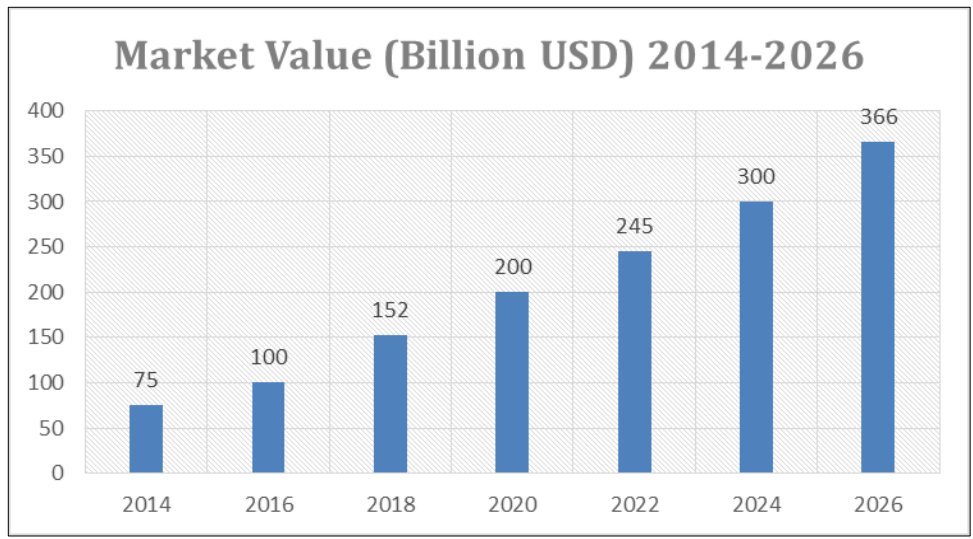

**Figure 6** Cybersecurity Investment Trends: CAGR Insights for 2014-2026

Figure 6. reflects the steady growth of the global cybersecurity market from **2014** to **2026**. Beginning at **$75 billion** in **2014**, the market expanded consistently due to rising cyber threats and technological advancements. By **2016**, it reached **$100 billion**, and this growth accelerated, hitting **$152 billion** in **2018** and **$200 billion** by **2020**. The upward trajectory continued with a market value of **$245 billion** in **2022**, driven by remote work and digitalization. Projected values for **2024** and **2026** are **$300 billion** and **$366 billion**, respectively, highlighting the increasing demand for robust cybersecurity solutions.

**4.3. Rising Trends in Financial and Cyber Fraud Cases: A Decade of Growth from 2014 to 2024:**

Table 5. [10], data on reported cyber fraud cases from 2014 to 2024, illustrates a troubling upward trend in various fraudulent activities. Financial fraud remains the most reported crime, escalating from 5,087 cases in 2014 to a projected 10,006 in 2024. This significant increase underscores the growing vulnerability of individuals and organizations to financial schemes. Phishing scams have also risen sharply, with reported cases climbing from 3,032 in 2014 to an anticipated 5,255 in 2024. These scams, which often deceive victims into revealing sensitive information, demonstrate the evolving tactics of cybercriminals. Online harassment cases increased from 2,021 in 2014 to 4,410 in 2024, highlighting the alarming growth of cyberbullying and other harmful online behaviors. Similarly, identity theft has seen a steady rise, with reports increasing from 1,528 in 2014 to 3,462 in 2024, revealing a growing concern over personal data security. Ransomware incidents have also become more frequent, rising from 826 cases in 2014 to an expected 1,748 in 2024. This trend indicates a significant challenge for cybersecurity, as these attacks typically disrupt operations and demand ransom payments. Overall, the data highlights the urgent need for enhanced cybersecurity measures, public awareness campaigns, and robust law enforcement strategies to combat the escalating threat of cybercrime in our increasingly digital world.

Here's the

**Table 5** Representing the number of fraud cases reported across various types of frauds from 2014 to 2024: [10]

| Year | Types of Fraud | Number of cases Reported |
|---|---|---|
| 2014 | Financial Fraud | 5,087 |
| 2014 | Phishing Scams | 3,032 |
| 2014 | Online Harassment | 2,021 |
| 2014 | Identity Theft | 1,528 |
| 2014 | Ransomware | 826 |
| 2014 | Credit Card Fraud | 1,231 |
| 2015 | Financial Fraud | 5,512 |

| 2015 | Phishing Scams | 3,211 |
|---|---|---|
| 2015 | Online Harassment | 2,567 |
| 2015 | Ransomware | 911 |
| 2015 | Credit Card Fraud | 1,308 |
| 2016 | Financial Fraud | 6,056 |
| 2016 | Phishing Scams | 3,507 |
| 2016 | Online Harassment | 2,700 |
| 2016 | Identity Theft | 1,801 |
| 2016 | Ransomware | 959 |
| 2016 | Credit Card Fraud | 1,401 |
| 2017 | Financial Fraud | 6,506 |
| 2017 | Phishing Scams | 3,709 |
| 2017 | Online Harassment | 3,030 |
| 2017 | Identity Theft | 2,020 |
| 2017 | Ransomware | 1,021 |
| 2017 | Credit Card Fraud | 1,523 |
| 2018 | Financial Fraud | 7,087 |
| 2018 | Phishing Scams | 4,021 |
| 2018 | Identity Theft | 2,223 |
| 2018 | Ransomware | 1,123 |
| 2018 | Credit Card Fraud | 1,610 |
| 2019 | Financial Fraud | 7,555 |
| 2019 | Phishing Scams | 4,205 |
| 2019 | Online Harassment | 3,412 |
| 2019 | Identity Theft | 2,405 |
| 2019 | Ransomware | 1,217 |
| 2019 | Credit Card Fraud | 1,729 |
| 2020 | Financial Fraud | 8,025 |
| 2020 | Phishing Scams | 4,454 |
| 2020 | Online Harassment | 3,622 |
| 2020 | Identity Theft | 2,614 |
| 2020 | Ransomware | 1,323 |
| 2020 | Credit Card Fraud | 1,841 |
| 2021 | Financial Fraud | 8,532 |
| 2021 | Phishing Scams | 4,656 |
| 2021 | Identity Theft | 2,844 |
| 2021 | Ransomware | 1,451 |
| 2021 | Credit Card Fraud | 1,961 |

| 2022 | Financial Fraud | 9,038 |
| 2022 | Phishing Scams | 4,846 |
| 2022 | Online Harassment | 4,015 |
| 2022 | Identity Theft | 3,041 |
| 2022 | Ransomware | 1,518 |
| 2022 | Credit Card Fraud | 2,049 |
| 2023 | Financial Fraud | 9,544 |
| 2023 | Phishing Scams | 5,062 |
| 2023 | Online Harassment | 4,236 |
| 2023 | Identity Theft | 3,222 |
| 2023 | Ransomware | 1,682 |
| 2023 | Credit Card Fraud | 2,174 |
| 2024* | Financial Fraud | 10,006 |
| 2024* | Phishing Scams | 5,255 |
| 2024* | Online Harassment | 4,410 |
| 2024* | Identity Theft | 3,462 |
| 2024* | Ransomware | 1,748 |
| 2024* | Credit Card Fraud | 2,233 |

From 2014 to 2024, the reported cases of various types of fraud have shown a significant upward trend. Financial fraud consistently topped the list, rising from 5,087 cases in 2014 to an anticipated 10,006 in 2024. Phishing scams also increased notably, growing from 3,032 cases in 2014 to 5,255 in 2024. Identity theft and online harassment cases have similarly surged, with identity theft cases rising from 1,528 to 3,462, and online harassment from 2,021 to 4,410. Ransomware incidents have experienced a gradual increase, reaching 1,748 in 2024, while credit card fraud cases have escalated from 1,231 to 2,233. Overall, the data reflects a concerning rise in cybercrimes, emphasizing the need for enhanced cybersecurity measures and public awareness to combat these persistent threats.

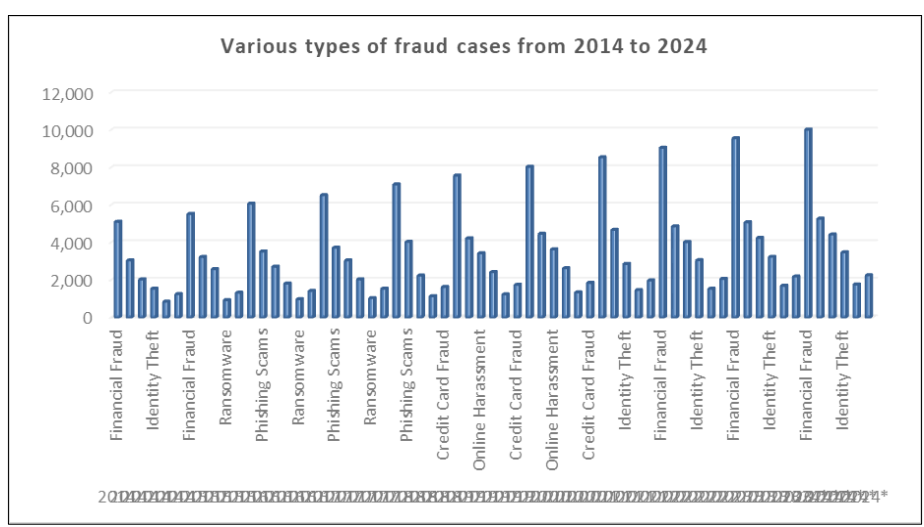

**Figure 7** Number of fraud cases reported across various types of frauds from 2014 to 2024

Figure 7 illustrates the trend of various types of fraud reported from 2014 to 2024, indicating a steady increase in cases over the years. Each year shows fluctuations in the number of reported incidents, but a clear upward trajectory is noticeable, particularly from 2020 onwards. Financial fraud consistently emerges as the most prevalent type, followed by phishing scams, online harassment, and identity theft. The spike in cases in 2024 suggests ongoing concerns about cybersecurity and the need for enhanced preventive measures. This data highlights the importance of awareness and proactive strategies to combat rising fraud rates in the digital age.

## 5. Conclusion

In conclusion, the accelerating integration of technology into daily life has markedly increased the risk of cybercrime, presenting significant challenges for individuals, businesses, and governments alike. This evolving digital landscape necessitates a prompt and robust enhancement of cybersecurity measures to protect sensitive data and digital assets. This paper highlights a comprehensive investigation of cybercrime, emphasizing the creation of novel prevention methods, enhancing internal security measures, and categorizing essential cybercrime terms to provide clearer insight into their impact on digital systems and infrastructure. The survey conducted indicates a notable rise in cybercrime incidents in India over the past decade, which emphasizes the urgent need for stronger legal frameworks that can keep pace with the evolving nature of cyber threats. A comprehensive public awareness campaign is crucial to educating citizens about the risks and protective measures associated with cybercrime. As highlighted by various studies, public engagement is fundamental in creating a culture of cybersecurity vigilance, wherein individuals are not just passive victims but active participants in safeguarding their digital lives. Moreover, our analysis of reported fraud cases across various categories from 2014 to 2024 reveals significant trends in financial fraud, phishing attacks, online harassment, identity theft, and ransomware incidents. Each of these categories reflects a shifting landscape of cybercrime, which has broader societal implications, from financial losses to emotional distress for victims. As the digital ecosystem continues to evolve, it is imperative that all stakeholders—government bodies, private organizations, and individuals—remain informed and coordinated in their responses to these challenges, ensuring a resilient framework for combating cybercrime in the future.